\documentclass{ifacconf}

\makeatletter
\let\old@ssect\@ssect 
\makeatother

\usepackage{natbib}								
\usepackage[nolist,nohyperlinks,printonlyused]{acronym}

\usepackage{booktabs}							
\usepackage{tabularx}							

\usepackage{supertabular}
\usepackage{multirow}

\usepackage{nicefrac}							
\usepackage{siunitx}							
\usepackage{amsmath,amssymb,amsfonts}
\usepackage{gensymb}
\usepackage{mathtools}							

\usepackage{graphicx}							
\usepackage{color}								
\usepackage{xcolor}
\usepackage{caption}
\usepackage{subcaption}
\usepackage{tikz}
\usepackage{pgfplots}
\usepackage{grffile}
\usepackage[european, americanvoltages, nooldvoltagedirection]{circuitikz}		

\usepackage[german,USenglish,UKenglish]{babel}
\usepackage[T1]{fontenc}
\usepackage{bm}									
\usepackage{dsfont}								
\usepackage{lmodern}							

\usepackage{xspace}
\usepackage{enumerate}							
\usepackage{cancel}								
\usepackage[normalem]{ulem}						
\usepackage{thmtools}							
\usepackage{hyphenat}							

\usepackage{xifthen}							

\usetikzlibrary{external}
\usetikzlibrary{decorations}
\usetikzlibrary{shapes,arrows}
\usetikzlibrary{calc}
\usetikzlibrary{angles}
\usetikzlibrary{positioning}
\usetikzlibrary{fit}
\usetikzlibrary{backgrounds}
\usetikzlibrary{mindmap,trees}

\pgfplotsset{compat=newest}
\usetikzlibrary{plotmarks}
\usetikzlibrary{arrows.meta}
\usetikzlibrary{bending}
\usetikzlibrary{shadows}
\usepgfplotslibrary{patchplots}

\usepackage{arydshln}					
\usepackage[colorlinks=true,linkcolor=black,anchorcolor=black,citecolor=black,filecolor=black,menucolor=black,runcolor=black,urlcolor=black]{hyperref}

\makeatletter
\def\@ssect#1#2#3#4#5#6{%
	\NR@gettitle{#6}
	\old@ssect{#1}{#2}{#3}{#4}{#5}{#6}
}
\makeatother

	\definecolor{purpleDark}{RGB}{118, 4, 205}
	\definecolor{purpleLight}{RGB}{186, 102, 250}
	
	\definecolor{blueDark}{RGB}{52, 78, 243}
	\definecolor{blueLight}{RGB}{118, 135, 244}
	\definecolor{redDark}{RGB}{197, 34, 0}
	\definecolor{redLight}{RGB}{255, 91, 57}
	\definecolor{yellowDark}{RGB}{255, 183, 0}
	\definecolor{yellowLight}{RGB}{255, 204, 77}
	\definecolor{greenDark}{RGB}{0, 143, 53}
	\definecolor{greenLight}{RGB}{42, 189, 97}
	
	\colorlet{greenFaint}{green!10!white}
	\colorlet{redFaint}{red!10!white}

	\definecolor{redText}{RGB}{222, 2, 10}
	\definecolor{orangeText}{RGB}{245, 86, 0}
	\definecolor{greenText}{RGB}{20,125,50}
	\definecolor{blueText}{RGB}{0, 114, 190}
	\definecolor{purpleText}{RGB}{115, 38, 146}
	\definecolor{pinkText}{RGB}{255, 107, 250}

	\definecolor{ballblue}{rgb}{0.13, 0.67, 0.8}
	\definecolor{buff}{rgb}{0.94, 0.86, 0.51}
	\definecolor{bronze}{rgb}{0.93,0.53,0.18}
	
	\definecolor{matlabCol1}{rgb}{0.0000,0.4470,0.7410}
	\definecolor{matlabCol2}{rgb}{0.8500,0.3250,0.0980}
	\definecolor{matlabCol3}{rgb}{0.9290,0.6940,0.1250}
	\definecolor{matlabCol4}{rgb}{0.4940,0.1840,0.5560}
	\definecolor{matlabCol5}{rgb}{0.4660,0.6740,0.1880}
	\definecolor{matlabCol6}{rgb}{0.3010,0.7450,0.9330}
	\definecolor{matlabCol7}{rgb}{0.6350,0.0780,0.1840}
	\definecolor{matlabCol8}{rgb}{0.0000,0.0000,0.0000}
	
	\newcommand{\autorefapp}[1]{\hyperref[#1]{Appendix~\ref*{#1}}}
	
	\addto\extrasUKenglish{%
	}

	\addto\extrasUSenglish{%
	}

	\makeatletter
	\newcommand\autorefMulti[1]{\@first@ref#1,@}
	\def\@throw@dot#1.#2@{#1}
	\def\@set@refname#1{
		\edef\@tmp{\getrefbykeydefault{#1}{anchor}{}}%
		\xdef\@tmp{\expandafter\@throw@dot\@tmp.@}%
		\ltx@IfUndefined{\@tmp autorefnameplural}%
			{\def\@refname{\@nameuse{\@tmp autorefname}s}}%
			{\def\@refname{\@nameuse{\@tmp autorefnameplural}}}%
	}
	\def\@first@ref#1,#2{%
		\ifx#2@\autoref{#1}\let\@nextref\@gobble
		\else%
			\@set@refname{#1}
			\@refname~\ref{#1}
			\let\@nextref\@next@ref
		\fi%
		\@nextref#2%
	}
	\def\@next@ref#1,#2{%
		\ifx#2@ and~\ref{#1}\let\@nextref\@gobble
		\else, \ref{#1}
		\fi%
		\@nextref#2%
	}
	\makeatother

	\newcommand{\Matlab}{{\rm \sc Matlab}\xspace}			
	\newcommand{\Simscape}{{\rm \sc Simscape}\xspace}		
	
	\DeclareSIUnit{\pu}{pu}							
	\DeclareSIUnit{\VAR}{\volt\ampere{}R}			
	
	\newcommand{\addWithPreComma}[1]{%
		\if\relax #1\relax
		\else%
		,#1
		\fi%
	}
	\newcommand{\addWithPostComma}[1]{%
		\if\relax #1\relax
		\else%
		#1,
		\fi%
	}
	\newcommand{\addInParentheses}[1]{%
	\if\relax #1\relax
	\else%
		(#1)
	\fi%
	}

\newacro{DDA}[DDA]{dynamics distributed averaging}
\newacro{DGU}[DGU]{distributed generation unit}

\newacro{P2G}[P2G]{power to gas}

\newacro{ISOPHS}[ISO-PHS]{input-state-output port-Hamiltonian system}
\newacro{PHS}[PHS]{port-Hamiltonian system}

\newacro{PDE}[PDE]{partial differential equation}
\newacro{ODE}[ODE]{ordinary differential equation}

\newacro{EIP}[EIP]{equilibrium-independent passive}
\newacro{IFP}[IFP]{input-feedforward passive}
\newacro{OFP}[OFP]{output-feedback passive}
\newacro{IFOFP}[IF-OFP]{input-feedforward output-feedback passive}

\newacro{ZSO}[ZSO]{zero-state observable}
\newacro{ZSD}[ZSD]{zero-state detectable}

\theoremstyle{plain}
\newtheorem{theorem}{Theorem}
\newtheorem{corollary}[theorem]{Corollary}

\newtheorem{assumption}{Assumption}
\newtheorem{proposition}[theorem]{Proposition}
\theoremstyle{definition}
\newtheorem{remark}{Remark}

	\renewcommand{\vec}[1]{\bm{#1}}						
	\renewcommand{\matrix}[1]{\bm{#1}}					

	\newcommand{\oNom}[1]{\tilde{#1}}					

	\newcommand{\Reals}{\ensuremath{\mathbb{R}}}		
	\newcommand{\RealsPos}{\ensuremath{\mathbb{R}_{+}}}	
	
	\newcommand{\vOneCol}[1][]{\vec{\mathds{1}}_{#1}}	
	\newcommand{\Ident}[1][]{\matrix{I}_{#1}}			
	
	\newcommand{\Transpose}{T}							
	\newcommand{\qedsymbol}{\hfill $\blacksquare$}		
	
	\DeclareMathOperator{\Diag}{Diag}					
	
	\newcommand{\dPartial}[2]{\dfrac{\partial #1}{\partial #2}}					
	
	\newlength\mytemplena
	\newlength\mytemplenb
	\DeclareDocumentCommand\myalignalign{sm}%
	{%
		\settowidth{\mytemplena}{$\displaystyle #2$}%
		\setlength\mytemplenb{\widthof{$\displaystyle=$}/2}%
		\hskip-\mytemplena%
		\hskip\IfBooleanTF#1{-\mytemplenb}{+\mytemplenb}%
	}

	\newcommand{\denoteCrit}{\mathrm{c}}
	\newcommand{\denoteEquivalent}{\mathrm{eq}}
	\newcommand{\denoteEffective}{\mathrm{e}}
	
	\newcommand{\denoteMean}{\mathrm{M}}
	\newcommand{\denoteSpecific}{\mathrm{s}}
	
	\newcommand{\denoteLaminar}{\mathrm{L}}
	\newcommand{\denoteTurbulent}{\mathrm{T}}
	
	\newcommand{\denoteLeft}{\mathrm{l}}
	\newcommand{\denoteMid}{\mathrm{m}}
	\newcommand{\denoteRight}{\mathrm{r}}

\newcommand{\sarea}[1][]{A_{#1}}

\newcommand{\mA}[1][]{\matrix{A}_{#1}}

\newcommand{\sbb}[1][]{b_{#1}}

\newcommand{\vb}[1][]{\vec{b}_{#1}}
\newcommand{\vbT}[1][]{\vec{b}_{#1}^\Transpose}

\newcommand{\mB}[1][]{\matrix{B}_{#1}}
\newcommand{\mBT}[1][]{\matrix{B}_{#1}^\Transpose}

\newcommand{\sbeta}[1][]{\beta_{#1}}
\newcommand{\sbetal}[1][]{\beta_{\denoteLeft\addWithPreComma{#1}}}
\newcommand{\sbetar}[1][]{\beta_{\denoteRight\addWithPreComma{#1}}}

\newcommand{\scspeed}[1][]{c_{#1}}
\newcommand{\scsqr}[1][]{\scspeed[#1]^2}

\newcommand{\sC}[1][]{C_{#1}}
\newcommand{\sCeq}[1][]{\sC[\denoteEquivalent\addWithPreComma{#1}]}
\newcommand{\sCeqinv}[1][]{\sC[\denoteEquivalent\addWithPreComma{#1}]^{-1}}

\newcommand{\sd}[1][]{d_{#1}}
\newcommand{\vd}[1][]{\vec{d}_{#1}}

\newcommand{\sdiam}[1][]{D_{#1}}

\newcommand{\ve}[1][]{\vec{e}_{#1}}
\newcommand{\veT}[1][]{\vec{e}_{#1}^\Transpose}

\newcommand{\mE}[1][]{\matrix{E}_{#1}}
\newcommand{\mET}[1][]{\mE[#1]^\Transpose}

\newcommand{\graphE}[1][]{\mathcal{E}_{#1}}

\newcommand{\svisc}[1][]{\eta_{#1}}

\newcommand{\sg}[1][]{g_{#1}}

\newcommand{\vg}[1][]{\vec{g}_{#1}}
\newcommand{\vgT}[1][]{\vec{g}_{#1}^\Transpose}

\newcommand{\mG}[1][]{\matrix{G}_{#1}}
\newcommand{\mGT}[1][]{\mG[#1]^\Transpose}

\newcommand{\graphG}[1][]{\mathcal{G}_{#1}}

\newcommand{\sfricEff}[1][]{\gamma_{#1}}

\newcommand{\sh}[1][]{h_{#1}}

\newcommand{\sH}[3]{H_{#1}^{#2}\addInParentheses{#3}}
\newcommand{\sHsx}[1][]{\sH{#1}{}{{\sx[#1]}}}
\newcommand{\sHx}[1][]{\sH{#1}{}{{\vx[#1]}}}

\newcommand{\sdHdx}[1][]{\dPartial{\sHsx[#1]}{\sx[#1]}}
\newcommand{\vdHdx}[1][]{\dPartial{\sHx[#1]}{\vx[#1]}}

\newcommand{\mJ}[1][]{\matrix{J}_{#1}}

\newcommand{\sk}[1][]{k_{#1}}
\newcommand{\skl}[1][]{\sk[\denoteLeft\addWithPreComma{#1}]}
\newcommand{\skr}[1][]{\sk[\denoteRight\addWithPreComma{#1}]}

\newcommand{\sK}[1][]{K_{#1}}

\newcommand{\sll}[1][]{l_{#1}}

\newcommand{\slen}[1][]{L_{#1}}

\newcommand{\seig}[1][]{\lambda_{#1}}

\newcommand{\sfric}[1][]{\lambda_{#1}}
\newcommand{\sfricL}[1][]{\sfric[\denoteLaminar\addWithPreComma{#1}]}
\newcommand{\sfricT}[1][]{\sfric[\denoteTurbulent\addWithPreComma{#1}]}
\newcommand{\sfrice}[1][]{\sfric[\denoteEffective\addWithPreComma{#1}]}

\newcommand{\sm}[1][]{m_{#1}}

\newcommand{\sM}[1][]{\dot{m}_{#1}}

\newcommand{\spp}[1][]{p_{#1}}
\newcommand{\spdot}[1][]{\dot{p}_{#1}}
\newcommand{\vp}[1][]{\vec{p}_{#1}}

\newcommand{\spc}[1][]{\spp[\denoteCrit\addWithPreComma{#1}]}
\newcommand{\spMean}[1][]{\spp[\denoteMean\addWithPreComma{#1}]}
\newcommand{\spn}[1][]{\oNom{p}_{#1}}

\newcommand{\spl}[1][]{\spp[\denoteLeft\addWithPreComma{#1}]}

\newcommand{\spr}[1][]{\spp[\denoteRight\addWithPreComma{#1}]}

\newcommand{\spldot}[1][]{\spdot[\denoteLeft\addWithPreComma{#1}]}
\newcommand{\sprdot}[1][]{\spdot[\denoteRight\addWithPreComma{#1}]}

\newcommand{\sphi}[1][]{\phi_{#1}}

\newcommand{\sq}[1][]{q_{#1}}
\newcommand{\sqabs}[1][]{|\sq[#1]|}

\newcommand{\sqn}[1][]{\oNom{q}_{#1}}
\newcommand{\sqndot}[1][]{\dot{\oNom{q}}_{#1}}
\newcommand{\sqnabs}[1][]{|\sqn[#1]|}
\newcommand{\vqn}[1][]{\oNom{\vec{q}}_{#1}}

\newcommand{\sqnl}[1][]{\sqn[\denoteLeft\addWithPreComma{#1}]}
\newcommand{\sqnm}[1][]{\sqn[\denoteMid\addWithPreComma{#1}]}
\newcommand{\sqnr}[1][]{\sqn[\denoteRight\addWithPreComma{#1}]}
\newcommand{\vqnm}[1][]{\vqn[\denoteMid\addWithPreComma{#1}]}

\newcommand{\sqnmdot}[1][]{\sqndot[\denoteMid\addWithPreComma{#1}]}
\newcommand{\sqnmabs}[1][]{\sqnabs[\denoteMid\addWithPreComma{#1}]}

\newcommand{\sQ}[1][]{Q_{#1}}

\newcommand{\mQ}[1][]{\matrix{Q}_{#1}}

\newcommand{\srough}[1][]{r_{#1}}

\newcommand{\sR}[1][]{R_{\denoteSpecific\addWithPreComma{#1}}}

\newcommand{\sRn}[1][]{\mathcal{R}_{#1}}
\newcommand{\sRnx}[1][]{\mathcal{R}_{#1}(\sx[#1])}

\newcommand{\mRnx}[1][]{\matrix{\mathcal{R}}_{#1}(\vx[#1])}

\newcommand{\sRe}[1][]{Re_{#1}}

\newcommand{\srho}[1][]{\rho_{#1}}
\newcommand{\srhon}[1][]{\oNom{\rho}_{#1}}

\newcommand{\st}[1][]{t_{#1}}

\newcommand{\sT}[1][]{T_{#1}}
\newcommand{\sTc}[1][]{\sT[\denoteCrit\addWithPreComma{#1}]}
\newcommand{\sTn}[1][]{\oNom{T}_{#1}}

\newcommand{\sinc}[1][]{\theta_{#1}}
\newcommand{\sinsinc}[1][]{\sin(\sinc[#1])}

\newcommand{\vu}[1][]{\vec{u}_{#1}}

\newcommand{\sv}[1][]{v_{#1}}
\newcommand{\svabs}[1][]{|v_{#1}|}

\newcommand{\sVol}[1][]{V_{#1}}

\newcommand{\graphV}[1][]{\mathcal{V}_{#1}}

\newcommand{\sx}[1][]{x_{#1}}
\newcommand{\sxdot}[1][]{\dot{x}_{#1}}
\newcommand{\vx}[1][]{\vec{x}_{#1}}
\newcommand{\vxT}[1][]{\vec{x}_{#1}^\Transpose}
\newcommand{\vxdot}[1][]{\dot{\vec{x}}_{#1}}

\newcommand{\vy}[1][]{\vec{y}_{#1}}

\newcommand{\sz}[1][]{z_{#1}}
\newcommand{\vz}[1][]{\vec{z}_{#1}}

\newcommand{\sZ}[1][]{Z_{#1}}

\graphicspath{{./03_Img/}}

\begin{document}
	\selectlanguage{UKenglish}
	
	\begin{frontmatter}
	
	\title{Port-Hamiltonian Modelling for Analysis and Control of Gas Networks\thanksref{footnoteinfo}} 
	
	\thanks[footnoteinfo]{This work was funded by Germany’s Federal Ministry for Economic Affairs and Climate Action (BMWK) as part of the RegEnZell project (reference number 0350062C).\\
	This work has been submitted to IFAC for possible publication.}
	
	\author[First]{Albertus J. Malan} 
	\author[First]{Lukas Rausche} 
	\author[First]{Felix Strehle}
	\author[First]{S{\"o}ren Hohmann}
	
	\address[First]{Institute of Control Systems (IRS),\\ Karlsruhe Institute of Technology (KIT), \\
		Kaiserstra{\ss}e 12, 76131 Karlsruhe, Germany (e-mail: \{albertus.malan,lukas.rausche,felix.strehle,soeren.hohmann\}@kit.edu).}

	\begin{abstract}
		
%
		In this paper, we present finite-dimensional \ac{PHS} models of a gas pipeline and a network comprising several pipelines for the purpose of control design and stability analysis. Starting from the partial differential Euler equations describing the dynamical flow of gas in a pipeline, the method of lines is employed to obtain a lumped-parameter model, which simplifies to a nonlinear third-order \ac{PHS}. 
		Parallels between gas networks and power systems are drawn by showing that the obtained pipeline \ac{PHS} model has the same $\pi$-representation as electrical transmission lines.
		Moreover, to assist future control design, additional passivity properties of the pipeline \ac{PHS} model are analysed and discussed.
		By comparing the proposed \ac{PHS} models against other models in a standard simulation, we show that the simplifying assumptions have no material effect on the model fidelity.
		The proposed pipeline and network models can serve as a basis for passivity-based control and analysis while the power system parallels facilitate the transfer of existing methods.
	\end{abstract}
	
	\acresetall
\end{frontmatter}

	\section{Introduction} \label{sec:Introduction}
%
The combination of \ac{P2G} facilities and the generation of green hydrogen envisions a sustainable and carbon-free future for gas networks. Since such \ac{P2G} and electrolysis facilities are ideally supplied by excess renewable energy, the supply of gas from such facilities are also subject to the intermittency and volatility associated with e.g.\ solar and wind power. Additionally, gas-fired electricity and heat generation is increasingly being used to compensate intermittent electrical energy generation, which can cause pressure fluctuations (see \cite{Osiadacz2020}). Due to the expected decrease in the overall demand for gas (see e.g.\ \cite{Qadrdan2019}), a \emph{coordination} of the \ac{P2G} facilities, local gas storages, flexible consumers and compressors supplying higher pressure networks will be required in the future.

For the control and subsequent stability analysis of such future gas networks, dynamical gas network models are required. Specifically, dynamical models are required for the pipelines, which are the most numerous components and typically exhibit the slowest dynamics. The flow of gas in the pipelines can accurately be described by the \acp{PDE} comprising the Euler equations. However, these \acp{PDE} pose a barrier to the application of many control and stability analysis methods applicable only to \acp{ODE}.
%
%
\paragraph*{Literature Review}
\cite{HerranGonzalez2009} and\linebreak \cite{Pambour2016} propose detailed simulation models for gas pipelines based on discretization. Similarly, \cite{Wiid2020} derive a nonlinear state-space model using the spectral element method for a pipeline with zero inclination. While the numerical nature of these models are appropriate for certain control methods (e.g.\ model predictive control), they generally provide no clear guidance towards control design or analytical system analysis. \cite{Ke2000} and \cite{Taherinejad2017} provide pipeline models for control design and analysis inspired by electrical analogies, although the assumptions made significantly affect the model fidelity in comparison with e.g.\ the simulation model in \cite{Pambour2016}. Furthermore, \cite{Alamian2012} propose a linearised state-space model and \cite{Zhou2017} presents linearised transfer functions for the pipeline dynamics. While the standard control and stability analysis methods can be applied to these linear models, the linearisation significantly impacts model accuracy. Finally, \cite{Domschke2021} propose various port-Hamiltonian-based models for a gas pipeline. \Ac{PHS} models may readily be used in passivity-based control and analysis methods without compromising model quality by removing nonlinear effects. Nevertheless, the infinite dimensional \ac{PHS} models presented by \cite{Domschke2021} retain the \ac{PDE} nature of the Euler equations, limiting their access to more general control and stability analysis methods in comparison to finite dimensional \acp{PHS}.
%
\paragraph*{Main Contribution}
In this paper, we propose finite dimensional \ac{PHS} models for a gas pipeline and a network of gas pipelines for which standard control and analysis techniques can be applied. Specifically, this comprises:
\begin{enumerate}
	\item A third order \ac{PHS} model with a $\pi$-model structure similar to those of power system transmission lines.
	\item A combined \ac{PHS} model for an entire network of pipes.
	\item A simulation demonstrating the fidelity of the proposed \ac{PHS} model compared to other models.
\end{enumerate}
Through parallels with models used for power systems and due to the use of the \ac{PHS} framework, the proposed models provide a gateway for transferring established control and analysis methods in the field of power systems to the domain of gas networks. Moreover, we further facilitate such a transfer of established methods by highlighting certain passivity properties of the \ac{PHS} models. 
%
\paragraph*{Paper Organisation}
The introduction concludes with some notation and preliminaries. In \autoref{sec:Prelim}, the equations representing the dynamics of an inclined gas pipeline are recalled. Next, in \autoref{sec:PHS_Model}, \ac{PHS} models are constructed for a gas pipeline and a network of pipelines. Thereafter in \autoref{sec:Simulation}, the fidelity of the proposed \ac{PHS} model is evaluated through a comparison with other pipeline models. Concluding remarks are supplied in \autoref{sec:Conclusion}.
%
\paragraph*{Notation and Preliminaries}
Define as a vector $\vec{a} = (a_i)$ and a matrix $\matrix{A} = (a_{ij})$. $\vOneCol[k]$ is a $k$-dimensional vector of ones and $\Ident[k]$ is the identity matrix of dimension $k$. $\Reals$ and $\RealsPos$ denote the real and positive real sets, respectively.
$\Diag[\cdot]$ creates a (block-)diagonal matrix from the supplied vectors (or matrices).
Note that we omit variable dependencies where clear from context.
%
We denote by $\graphG = (\graphV, \graphE)$ a finite, undirected graph with vertices $\graphV$ and edges $\graphE \subseteq \graphV \times \graphV$. Let $|\graphV|$ be the cardinality of the set $\graphV$.
By arbitrarily assigning directions to each edge in $\graphE$, the \emph{incidence matrix} $\mB \in \Reals^{|\graphV|\times|\graphE|}$ of $\graphG$ is defined by
\begin{equation} \label{eq:Intro:Incidence}
	\sbb[ij] = \left\{\begin{array}{@{}rl}
		+1 & \text{if vertex $i$ is the sink of edge $j$}, \\
		-1 & \text{if vertex $i$ is the source of edge $j$}, \\
		0 & \text{otherwise}.
	\end{array}\right.
\end{equation}

	\section{Gas Pipeline Preliminaries} \label{sec:Prelim}
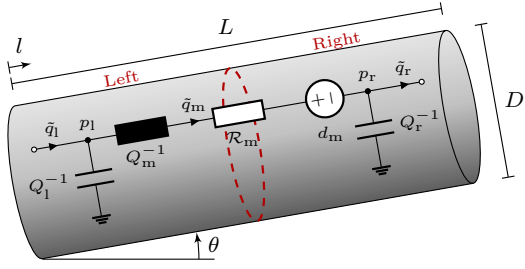
\begin{figure}[!t]
	\centering
	\resizebox{0.8\columnwidth}{!}{%
		\tikzsetnextfilename{03_Img/gas_pipeline}%
\begin{tikzpicture}
	\def\nodeXdist{5cm}
	\def\nodeYdist{2.4cm}
	\def\nodeXshift{\nodeXdist/2}
	
	\def\rotCylinder{10}
	\def\widthCylinder{2cm}
	\def\lengthCylinder{3*\widthCylinder}
	
	\def\busGendist{20pt}
	
	\node[cylinder, black, rotate=\rotCylinder, draw, minimum height=\lengthCylinder, minimum width=\widthCylinder, cylinder end fill=gray!50, top color=gray!0, bottom color=gray!20!black, middle color=gray!50, shading angle=\rotCylinder, aspect=1.75] (nCylinder) {};
	\path (nCylinder.after bottom) -- (nCylinder.before top) coordinate[pos=0.505] (cCut);
	\draw[dashed, rotate=\rotCylinder-90, red!70!black, thick] (cCut) ++(\rotCylinder-90:1pt) arc (180:540:1cm and 0.2cm);
	\path (nCylinder.after bottom) -- (nCylinder.before top) node[pos=0.25,sloped,above=-2.5pt,red!70!black] {\tiny Left} node[pos=0.75,sloped,above=-3.5pt,red!70!black] {\tiny Right};
	
	\path (nCylinder.after bottom) ++(\rotCylinder+90:8pt) coordinate (cSpatialStart);
	\path (nCylinder.before top) ++(\rotCylinder+90:8pt) coordinate (cSpatialEnd);
	\path (cSpatialStart) ++(\rotCylinder+90:6pt) coordinate (cLengthStart) ++ (\rotCylinder:0.3cm) coordinate (cLengthEnd);
	
	\draw[{Bar[]}-{Bar[]}] (cSpatialStart) -- (cSpatialEnd) node[midway,above,anchor=south] {\small $\slen$};
	\draw[{Bar[]}-latex'] (cLengthStart) -- (cLengthEnd) node[midway,above,anchor=south] {\small $\sll$};
	
	\path (nCylinder.before top) ++(\rotCylinder:13pt) coordinate (cDiameterStart);
	\path (nCylinder.after top) ++(\rotCylinder:13pt) coordinate (cDiameterEnd);
	
	\draw[{Bar[]}-{Bar[]}] (cDiameterStart) -- (cDiameterEnd) node[midway,anchor=west] {\small $\sdiam$};
	
	\path (nCylinder.before bottom) ++(0:2.2cm) coordinate (cAngleBase);
	\path (nCylinder.before bottom) ++(0:2cm) coordinate (cAngleStart);
	\path (nCylinder.before bottom) ++(\rotCylinder:2cm) coordinate (cAngleEnd);
	
	\draw (nCylinder.before bottom) -- (cAngleBase);
	\draw[-latex'] (cAngleStart) arc [start angle=0, end angle=\rotCylinder, radius=2cm] node[midway,anchor=west] {\small $\sinc$};
	
	\path (nCylinder.after bottom) -- (nCylinder.before bottom) coordinate[pos=0.3] (cCircuitL);
	\path (nCylinder.before top) -- (nCylinder.after top) coordinate[pos=0.3] (cCircuitR);
	\path (cCircuitL) -- (cCircuitR)
		coordinate[pos=0.025] (cPortL)
		coordinate[pos=0.925] (cPortR)
		coordinate[pos=0.15] (cCapLTop)
		coordinate[pos=0.8] (cCapRTop)
		coordinate[pos=0.15] (cIndStart)
		coordinate[pos=0.4] (cIndEnd)
		coordinate[pos=0.4] (cResStart)
		coordinate[pos=0.6] (cResEnd)
		coordinate[pos=0.6] (cGravStart)
		coordinate[pos=0.8] (cGravEnd);
	\path (cCapLTop) ++(\rotCylinder-90:0.5*\widthCylinder) coordinate (cCapLBot);
	\path (cCapRTop) ++(\rotCylinder-90:0.5*\widthCylinder) coordinate (cCapRBot);
	
	\ctikzset{bipoles/length=.8cm}
	\ctikzset{bipole annotation style/.style={font=\tiny}}
	\ctikzset{bipole current style/.style={font=\tiny}}
	\ctikzset{bipoles/vsourceam/inner plus={\tiny$+$}}
	\ctikzset{bipoles/vsourceam/inner minus={\tiny$-$}}
	\ctikzset{current/distance = 0}
	
	\draw (cIndStart) to [open, i=$ $, name=CurrM] (cResEnd);
	\ctikzset{current/distance = -1}
	\draw (cPortL) to [open, i=$ $, name=CurrL] (cCapLTop);
	\ctikzset{current/distance = 1}
	\draw (cPortR) to [open, i<=$ $, name=CurrR] (cCapRTop);
	
	\draw (cPortL) to [short, o-] 
		(cIndStart) to [L, a=${\sQ[\denoteMid]^{-1}}$] (cIndEnd) to [-]
		(cResStart) to [R, fill=white, a=${\sRn[\denoteMid]}$] (cResEnd) to [-]
		(cGravStart) to [american voltage source, fill=white, a=${\sd[\denoteMid]}$] (cGravEnd) to [short,-o]
		(cPortR);
	\draw (cCapLTop) to [C, *-, a=${\sQ[\denoteLeft]^{-1}}$] (cCapLBot) node[tlground, rotate=\rotCylinder]{};
	\draw (cCapRTop) to [C, *-, a^=${\sQ[\denoteRight]^{-1}}$] (cCapRBot) node[tlground, rotate=\rotCylinder]{};
	
	\node[anchor=south] at(CurrM-Ipos) {\tiny$\sqnm$};
	\node[anchor=south] at(CurrL-Ipos) {\tiny$\sqnl$};
	\node[anchor=south] at(CurrR-Ipos) {\tiny$\sqnr$};
	
	\node[anchor=south] at(cCapLTop) {\tiny$\spl$};
	\node[anchor=south] at(cCapRTop) {\tiny$\spr$};

\end{tikzpicture}
%
	}
	\caption{A gas pipeline sectioned into a left and a right side, superimposed with its electrical equivalent circuit.}
	\label{fig:Prelim:Pipeline}
\end{figure}
In this section, we briefly recall the equations governing the flow of gas in an inclined pipeline (see \autoref{fig:Prelim:Pipeline}). We refer the reader to \cite{Koch2015,Pambour2016} and the sources therein for a derivation of the equations presented in this section. 

Recall that the Euler equations for the flow dynamics of the pipeline in \autoref{fig:Prelim:Pipeline} under isothermal conditions are described by the \acp{PDE}
\begingroup
\allowdisplaybreaks
\begin{align}
	\label{eq:Prelim:Pipe_mass_conserv}
	\dPartial{\srho}{\st} &+ \dPartial{(\srho \sv)}{\sll} = 0, \\
	\label{eq:Prelim:Pipe_momentum_conserv}
	\dPartial{(\srho \sv)}{\st} &+ \dPartial{(\srho \sv^2)}{\sll} + \dPartial{\spp}{\sll} + \frac{\sfrice \srho \svabs \sv}{2 \sdiam} + \srho \sg \sinsinc = 0,
\end{align}
\endgroup
which comprise equations for the conservation of mass \eqref{eq:Prelim:Pipe_mass_conserv} and the conservation of momentum \eqref{eq:Prelim:Pipe_momentum_conserv}. For a pipe with length $\slen$ and diameter $\sdiam$, $\sll \in [0,\slen]$ is the spatial variable, $\sv = \sv(\sll,\st) \in \Reals$ is the gas velocity, $\srho=\srho(\sll,\st) > 0$ is the density, $\spp=\spp(\sll,\st) > 0$ is the pressure, $\sfrice = \sfrice(\sv,\srho) > 0$ is the effective friction factor, $\sg$ is the gravitational acceleration, and $\sinc \in [-\nicefrac{\pi}{2},\nicefrac{\pi}{2}]$ is the pipe inclination. We also recall the real gas law
\begin{equation} \label{eq:Prelim:Real_gas_law_isothermal}
	\spp = \sZ \sR \sT \srho = \scsqr \srho ,
\end{equation}
where $\sR$ is the specific gas constant, $\sT$ the temperature, $\scspeed$ is the speed of sound and $\sZ = \sZ(\spp, \sT) > 0$ is the compressibility factor for real gasses. For natural gas up to $\SI{150}{\bar}$, $\sZ$ can be estimated using the \emph{Papay} approximation
\begin{equation} \label{eq:Prelim:Gas_compressibility}
	\sZ(\spp, \sT) = 1 - 3.52\frac{\spp}{\spc} e^{-2.26\frac{\sT}{\sTc}} + 0.274\frac{\spp^2}{\spc^2} e^{-1.878\frac{\sT}{\sTc}},
\end{equation}
where $\sTc$ and $\spc$ are the critical temperature and critical pressure, respectively.

The friction factor $\sfric$ depends on the state of flow (laminar, turbulent or somewhere in between), which is described by the \emph{Reynolds} number
\begin{equation} \label{eq:Prelim:Reynolds}
	\sRe = \frac{\srho \svabs \sdiam}{\svisc} = \frac{\srho \sqabs \sdiam}{\svisc \sarea} ,
\end{equation}
with the dynamic viscosity $\svisc$, the cross-sectional area $\sarea = {\pi\sdiam^2}/{4}$, and the volumetric flow rate $\sq$.
Friction under laminar flow conditions, typically with $\sRe < 2300$, is characterised by the \emph{Hagen-Poisseuille} formula
\begin{equation} \label{eq:Prelim:Friction_laminar}
	\sfricL = \frac{64}{\sRe},
\end{equation}
whereas the friction factor for turbulent flow, typically with $\sRe \ge 2300$, is described using the implicit \emph{Colebrook-White} equation
\begin{equation} \label{eq:Prelim:Friction_turbulent_implicit}
	\frac{1}{\sqrt{\sfricT}} = -2 \log_{10} \left(\frac{2.51}{\sRe \sqrt{\sfricT}} + \frac{\srough}{3.71 \sdiam}\right) ,
\end{equation}
where $\srough$ is the surface roughness. The implicit form \eqref{eq:Prelim:Friction_turbulent_implicit} can be approximated in gas pipelines using the \emph{Hofer} equation
\begin{equation} \label{eq:Prelim:Friction_turbulent_hofer}
	\sfricT = \left[2 \log_{10} \left(\frac{4.518}{\sRe} \log_{10} \left(\frac{\sRe}{7}\right) + \frac{\srough}{3.71 \sdiam} \right)\right]^{-2}.
\end{equation}
Moreover, a friction efficiency factor $\sfricEff$ can be included to account for changes in the pipe curvature or form, where
\begin{equation} \label{eq:Prelim:Friction_efficiency}
	\sqrt{\frac{1}{\sfrice}} = \sfricEff \sqrt{\frac{1}{\sfric}} \iff \sfrice = \frac{\sfric}{\sfricEff^2} ,
\end{equation}
is the effective friction factor and where $\sfric$ is set to $\sfricT$ \eqref{eq:Prelim:Friction_turbulent_hofer} or $\sfricL$ \eqref{eq:Prelim:Friction_laminar}, depending on $\sRe$.

The pipeline \acp{PDE} in \eqref{eq:Prelim:Pipe_mass_conserv} and \eqref{eq:Prelim:Pipe_momentum_conserv} can be simplified by assuming slow velocities and by converting to standard conditions. For slow velocities ($|\sv| \le \SI{15}{\meter\per\second}$), \cite{Pambour2016} observe that\footnote{Taking $\scspeed \approx \SI{300}{\meter\per\second}$, we find that $\sv^2/\scsqr < \num{2.5e-3} \ll 1$.}
\begin{equation} \label{eq:Prelim:spatial_simplification}
	\begin{aligned}
		\dPartial{(\srho \sv^2)}{\sll} + \dPartial{\spp}{\sll} &= \dPartial{}{\sll}\left[\spp\left(\frac{\sv^2}{\scsqr} + 1 \right)\right] 
		\approx \dPartial{\spp}{\sll} .
	\end{aligned}
\end{equation}
Additionally, the mass flow rate $\sM$ relates the volumetric flow rate $\sq$ and the velocity $\sv$ to the \emph{volumetric flow rate at standard conditions} $\sqn$ according to 
\begin{equation} \label{eq:Prelim:Flow_rate_relation}
	\sM = \srho \sarea \sv = \srho \sq = \srhon \sqn
\end{equation}
where $\srhon$ is the constant standard density. Through \eqref{eq:Prelim:Real_gas_law_isothermal}, \eqref{eq:Prelim:spatial_simplification} and \eqref{eq:Prelim:Flow_rate_relation}, the \acp{PDE} in \eqref{eq:Prelim:Pipe_mass_conserv} and \eqref{eq:Prelim:Pipe_momentum_conserv} simplify to
\begingroup
\allowdisplaybreaks
\begin{align}
	\label{eq:Prelim:Simplified_mass_conserv}
	\frac{1}{\scsqr} \dPartial{\spp}{\st} &= - \frac{\srhon}{\sarea} \dPartial{\sqn}{\sll}, \\
	\label{eq:Prelim:Simplified_momentum}
	\frac{\srhon}{\sarea} \dPartial{\sqn}{\st} &= - \dPartial{\spp}{\sll} - \frac{\sfrice \srhon^2 \scsqr}{2 \sfricEff^2 \sdiam \sarea^2 \spp} \sqnabs \sqn - \frac{\sg \sinsinc}{\scsqr} \spp.
\end{align}
\endgroup
%
%

	\section{Port-Hamiltonian Modelling} \label{sec:PHS_Model}
Building on the simplified mass and momentum equations, \eqref{eq:Prelim:Simplified_mass_conserv} and \eqref{eq:Prelim:Simplified_momentum}, we now construct an \ac{ISOPHS} model for the gas pipeline. For simplicity, we refer to the \acp{ISOPHS} simply as \ac{PHS} models. In \autoref{sec:PHS_Mode:Lumped_Param}, we start by deriving a lumped-parameter model using the method of lines. Thereafter in \autoref{sec:PHS_Mode:PHS}, the obtained differential equations are combined into a single \ac{PHS} representing the pipeline. Finally, in \autoref{sec:PHS_Model:Network}, a \ac{PHS} model is proposed for a network of gas pipelines.
\subsection{Lumped-Parameter Model} \label{sec:PHS_Mode:Lumped_Param}
Recall that the method of lines allows \acp{PDE} to be converted to \acp{ODE} through discretization of the partial derivative terms (see e.g.\ \cite{Cellier2006}). We therefore employ the spatial approximations
\begin{align}
	\label{eq:PHS_Model:Approx_qn}
	\left.\dPartial{\sqn}{\sll} \right|_{1,2} &\approx \frac{\sqn[2] - \sqn[1]}{\varDelta \sll} \\
	\label{eq:PHS_Model:Approx_p}
	\left.\dPartial{\spp}{\sll} \right|_{1,2} &\approx \frac{\spp[2] - \spp[1]}{\varDelta \sll}
\end{align}
between two arbitrary points on the pipeline. For the lumped-parameter model, we divide the pipeline into left ($\denoteLeft$) and right ($\denoteRight$) sides which connect in the middle ($\denoteMid$) as in \autoref{fig:Prelim:Pipeline}.
\begin{proposition} \label{prop:PHS_Model:Mass_ODE}
	The simplified mass flow \ac{PDE} \eqref{eq:Prelim:Simplified_mass_conserv} for a pipeline can be represented by the \acp{ODE}
	\begin{equation} \label{eq:PHS_Model:Mass_ODE}
		\left\{
		\begin{aligned}
			\spldot &= \frac{2 \srhon \scsqr}{\slen \sarea} (\sqnl - \sqnm), \\
			\sprdot &= \frac{2 \srhon \scsqr}{\slen \sarea} (\sqnm - \sqnr).
		\end{aligned}
		\right.
	\end{equation}
\end{proposition}
\begin{pf}
	Divide the pipeline into two sections as in \autoref{fig:Prelim:Pipeline} such that the discretization in \eqref{eq:PHS_Model:Approx_p} yields
	\begin{equation} \label{eq:PHS_Model:Discrete_qn}
		\left.\dPartial{\sqn}{\sll} \right|_{\denoteLeft,\denoteMid} \approx \frac{2}{\slen} (\sqnm - \sqnl), \quad \left.\dPartial{\sqn}{\sll} \right|_{\denoteMid,\denoteRight} \approx \frac{2}{\slen} (\sqnr - \sqnm)
	\end{equation}
	for the left and the right sections, respectively. Substitute \eqref{eq:PHS_Model:Discrete_qn} into \eqref{eq:Prelim:Simplified_mass_conserv} to obtain the \acp{ODE} in \eqref{eq:PHS_Model:Mass_ODE}. \qedsymbol
\end{pf}
\begin{proposition} \label{prop:PHS_Model:Momentum_ODE}
	The simplified momentum \ac{PDE} \eqref{eq:Prelim:Simplified_momentum} for a pipeline can be represented by the \ac{ODE}
	\begin{equation} \label{eq:PHS_Model:Momentum_ODE}
		\frac{\srhon}{\sarea} \sqnmdot = \frac{\spl - \spr}{\slen} - \frac{\sfrice \srhon^2 \scsqr \sqnmabs}{2 \sfricEff^2 \sdiam \sarea^2 \spMean} \sqnm - \frac{\sg \sinsinc}{\scsqr} \spMean ,
	\end{equation}
	where $\spMean$ is the mean pressure in the pipeline with
	\begin{equation} \label{eq:PHS_Model:Mean_Pressure}
		\spMean = \frac{2}{3} \frac{\spl^3 - \spr^3}{\spl^2 - \spr^2} = \frac{2}{3} \left(\spl + \spr - \frac{\spl \cdot \spr}{\spl + \spr}\right) .
	\end{equation}
\end{proposition}
\begin{pf}
	Consider the discretization for the entire length of the pipeline in \autoref{fig:Prelim:Pipeline} and replace the differential pressure term ${\partial \spp}/{\partial \sll}$ in \eqref{eq:Prelim:Simplified_momentum} with \eqref{eq:PHS_Model:Approx_p}, where $\varDelta\sll = \slen$. Furthermore, replace the pressure terms in \eqref{eq:Prelim:Simplified_momentum} with the average pressure in the pipeline \eqref{eq:PHS_Model:Mean_Pressure} to obtain \eqref{eq:PHS_Model:Momentum_ODE}.
	\qedsymbol
\end{pf}
\begin{remark}[Mean pressure] \label{rem:PHS_Model:Mean_Pressure}
	The mean pressure in a\break pipeline \eqref{eq:PHS_Model:Mean_Pressure} was originally derived by \cite[p.~203]{Weymouth1912} in his calculation for the volume of gas in a pipe. A proof is also given in \cite[Lemma~2.3]{Koch2015}.
\end{remark}
\begin{remark}[Discretization choice] \label{rem:PHS_Model:Equivalent_discretization}
	The chosen discretization in \autorefMulti{prop:PHS_Model:Mass_ODE, prop:PHS_Model:Momentum_ODE} is similar in principle to the approach in \cite{Pambour2016}, where it is applied to static system equations. Note that higher order pipeline models with corresponding \ac{PHS} representations may be generated by increasing the number of discretization points in the same scheme as presented here.
\end{remark}
\subsection{Port-Hamiltonian Representation} \label{sec:PHS_Mode:PHS}
Using the \acp{ODE} provided by \autorefMulti{prop:PHS_Model:Mass_ODE, prop:PHS_Model:Momentum_ODE}, we now proceed with constructing a single \ac{PHS} for the dynamical gas pipeline. To allow for the \ac{PHS} formulation, we make the following additional simplifying assumptions, the validity of which is discussed in the sequel.
\begin{assumption} \label{asm:PHS_Model:Constant_values}
	The compressibility $\sZ$ in \eqref{eq:Prelim:Gas_compressibility} is constant\footnote{As an example, analysing \eqref{eq:Prelim:Gas_compressibility} at $\SI{0}{\degreeCelsius}$ shows that $\sZ$ decreases with a slope no larger than $\SI{0.3}{\percent\per\bar}$.}. 
\end{assumption}
\begin{assumption} \label{asm:PHS_Model:Positive_pressure}
	The pressure in the pipe is always positive.
\end{assumption}
\begin{assumption} \label{asm:PHS_Model:Constant_graviational_effect}
	The effect of the height difference in \eqref{eq:PHS_Model:Momentum_ODE}, i.e.\ the last term in \eqref{eq:PHS_Model:Momentum_ODE} and specifically $\spMean$, is constant.
\end{assumption}
\begin{theorem}[Pipeline \ac{PHS} model] \label{thm:PHS_Model:PHS_Pipe}
	Let \autorefMulti{asm:PHS_Model:Constant_values, asm:PHS_Model:Positive_pressure, asm:PHS_Model:Constant_graviational_effect} hold. Then, the dynamics of the isothermal gas pipeline in \autoref{fig:Prelim:Pipeline} can be written as the following \ac{PHS}:
	\begin{subequations} \label{eq:PHS_Model:Nonlinear_PHS_full}
		\begin{equation} \label{eq:PHS_Model:Nonlinear_PHS}
			\left\lbrace
			\begin{aligned}
				\vxdot &= (\mJ - \mRnx)\vdHdx + \mG \vu + \ve \sd, \\
				\vy &= \mGT \vdHdx , \quad
				\sz = \veT \vdHdx , \\
				\sHx &= \frac{1}{2} \vxT \mQ \vx,
			\end{aligned}
			\right.
		\end{equation}
		with states \eqref{eq:PHS_Model:PHS_states}, co-states \eqref{eq:PHS_Model:PHS_co_states}, input-output port pair \eqref{eq:PHS_Model:PHS_inputs_outputs}, and disturbance port pair \eqref{eq:PHS_Model:PHS_disturbance}
		\begingroup
		\allowdisplaybreaks
		\begin{align}
			\label{eq:PHS_Model:PHS_states}
			\vx &= \begin{bmatrix}
				\dfrac{\slen \sarea}{2 \srhon \scsqr} \spl \;\;& \dfrac{\slen \sarea}{2 \srhon \scsqr} \spr \;\;& \dfrac{\srhon \slen}{\sarea} \sqnm
			\end{bmatrix}^\Transpose, \\
			\label{eq:PHS_Model:PHS_co_states}
			\vdHdx &= \mQ\vx = \begin{bmatrix}
				\spl & \spr & \sqnm
			\end{bmatrix}^\Transpose,\\
			\label{eq:PHS_Model:PHS_inputs_outputs}
			\vu &= \begin{bmatrix}
				\sqnl & -\sqnr
			\end{bmatrix}^\Transpose,
			\qquad
			\vy = \begin{bmatrix}
				\spl & \spr
			\end{bmatrix}^\Transpose,\\
			\label{eq:PHS_Model:PHS_disturbance}
			\sd &= \frac{\sg \slen \sinsinc}{\scsqr} \spMean,
			\qquad
			\sz = \sqnm ,
		\end{align}
		\endgroup
		where $\sd$ is constant and with the interconnection structure \eqref{eq:PHS_Model:J_matrix}, resistive structure \eqref{eq:PHS_Model:R_Matrix}, input matrix \eqref{eq:PHS_Model:Input_matrix}, disturbance matrix \eqref{eq:PHS_Model:Disturbance_matrix} and storage matrix \eqref{eq:PHS_Model:Weight_matrix}
		\begingroup
		\allowdisplaybreaks
		\begin{align}
			\label{eq:PHS_Model:J_matrix}
			\mJ &= \begin{bmatrix}
				0 & 0 & -1 \\
				0 & 0 & 1 \\
				1 & -1 & 0
			\end{bmatrix},\\
			\label{eq:PHS_Model:R_Matrix}
			\mRnx &= \Diag\left[0, \; 0, \; \frac{\sfrice \srhon^2 \scsqr \slen \sqnmabs}{2 \sdiam \sarea^2 \spMean}\right],\\
			\label{eq:PHS_Model:Input_matrix}
			\mGT &= \begin{bmatrix}
				1 & 0 & 0 \\
				0 & 1 & 0
			\end{bmatrix},\\
			\label{eq:PHS_Model:Disturbance_matrix}
			\veT &= \begin{bmatrix}
				0 & 0 & -1
			\end{bmatrix},\\
			\label{eq:PHS_Model:Weight_matrix}
			\mQ &= \Diag\left[ \frac{2 \srhon \scsqr}{\slen \sarea}, \; \frac{2 \srhon \scsqr}{\slen \sarea}, \; \frac{\sarea}{\srhon \slen} \right].
		\end{align}
		\endgroup
	\end{subequations}
\end{theorem}
\begin{pf}
	The \ac{PHS} in \eqref{eq:PHS_Model:Nonlinear_PHS_full} is constructed by combining \eqref{eq:PHS_Model:Mass_ODE} and \eqref{eq:PHS_Model:Momentum_ODE} and rearranging the terms to obtain $\vxdot$ on the left-hand side, where desired states $\vx$ are given by \eqref{eq:PHS_Model:PHS_states}. Note that $\sd$ is constant due to \autoref{asm:PHS_Model:Constant_graviational_effect}. To show that \eqref{eq:PHS_Model:Nonlinear_PHS_full} is a \ac{PHS}, we verify that $\mQ$ is positive definite and constant by \autoref{asm:PHS_Model:Constant_values}, since $\scspeed$ depends on $\sZ$, and hence that $\sH{}{}{}$ is positive definite in $\vx$. Lastly, we verify that $\mRnx \ge 0$ since $\sfrice > 0$ for all $\sqnm \in \Reals$ and by invoking \autoref{asm:PHS_Model:Positive_pressure}, i.e.\ $\spMean > 0$. \qedsymbol
\end{pf}
Through \autoref{thm:PHS_Model:PHS_Pipe}, the gas pipeline is presented in a manner that readily allows passivity-based analysis and control methods to be applied. Moreover, as a direct result of the discretization choices in \autoref{sec:PHS_Mode:Lumped_Param}, the presented \ac{PHS} bears a strong resemblance to the $\pi$-models used for the transmission lines in electrical networks (see \autoref{fig:Prelim:Pipeline}, c.f.\ \cite{Strehle2020DC}). The gas pipeline model in \autoref{thm:PHS_Model:PHS_Pipe} deviates from the electrical $\pi$-lines only due to the nonlinearity arising from the resistive structure and the disturbance term due to the height difference.

In the case of a network comprising electrical $\pi$-model transmission lines, the capacitive legs of the $\pi$-model are often separated from the inductive-resistive components. This allows the capacitive effects from various lines to be combined into a single component at a network node and yields simplified line dynamics without compromising the model accuracy (see e.g.\ \cite{Strehle2020DC}). While this idea has also been applied to static gas pipelines models (see e.g. \cite{Pambour2016}), the following corollary derived from \autoref{thm:PHS_Model:PHS_Pipe} considers this capacitive separation in the \ac{PHS} framework.
\begin{corollary}[Split pipeline \acp{PHS}] \label{thm:PHS_Model:Split_PHS}
	The \ac{PHS} gas pipeline model in \autoref{thm:PHS_Model:PHS_Pipe}, where \autorefMulti{asm:PHS_Model:Constant_values, asm:PHS_Model:Positive_pressure, asm:PHS_Model:Constant_graviational_effect} hold, is equivalent to the combination of the inductive-resistive dynamics between the left and right sides
	\begin{equation} \label{eq:PHS_Model:Split_RL}
		\left\lbrace
		\begin{aligned}
			\underbrace{\frac{\srhon \slen}{\sarea} \sqnmdot}_{\textstyle \sxdot[\denoteMid]} &= {-} \underbrace{\!\frac{\sfrice \srhon^2 \scsqr \slen \sqnmabs}{2 \sdiam \sarea^2 \spMean}}_{\textstyle \sRnx[\denoteMid]} \sqnm {+} \underbrace{\begin{bmatrix} 1 & -1 \end{bmatrix}}_{\textstyle \vgT[\denoteMid]} \underbrace{\begin{bmatrix} \spl \\ \spr \end{bmatrix}}_{\textstyle \vu[\denoteMid]} {-} \underbrace{\frac{\sg \slen \! \sinsinc}{\scsqr} \spMean}_{\textstyle \sd[\denoteMid]} , \\
			\vy[\denoteMid] &= \vg[\denoteMid] \sdHdx[\denoteMid] = \begin{bmatrix} \sqnm \\ -\sqnm \end{bmatrix}, \\
			\sz[\denoteMid] &= \sdHdx[\denoteMid] = \sqnm, \\
			\sx[\denoteMid] &= \frac{\srhon \slen}{\sarea} \sqnm, \quad \sQ[\denoteMid] = \frac{\sarea}{\srhon \slen}, \quad \sHsx[\denoteMid] = \frac{1}{2} \sQ[\denoteMid] \sx[\denoteMid]^2,
		\end{aligned}
		\right.
	\end{equation}
	and the capacitive dynamics at the left and right side of the pipeline
	\begin{equation} \label{eq:PHS_Model:Split_C}
		\left\lbrace
		\begin{aligned}
			\underbrace{\frac{\slen \sarea}{2 \srhon \scsqr} \spdot[k]}_{\textstyle \sxdot[k]} &= \underbrace{\begin{bmatrix} 1 & 1 \end{bmatrix}}_{\textstyle \vgT[k]} \underbrace{\begin{bmatrix} \sbeta[k]\sqn[k] \\ -\sbeta[k]\sqnm \end{bmatrix}}_{\textstyle \vu}, \\
			\vy[k] &= \vg[k] \sdHdx[k] = \begin{bmatrix} \spp[k] \\ \spp[k] \end{bmatrix}, \\
			\sHsx[k] &= \frac{1}{2} \sQ[k] \sx[k]^2, \quad \sQ[k] = \frac{2 \srhon \scsqr}{\slen \sarea}, \quad \sx[k] = \frac{\slen \sarea}{2 \srhon \scsqr} \spp[k],
		\end{aligned}
		\right.
	\end{equation}
	where $k \in \{\denoteLeft, \denoteRight\}$, $\sbetal = 1$ and $\sbetar = -1$.
\end{corollary}
\begin{pf}
	By using the states $\spl$ and $\spr$ as coupling inputs for the first two states in \eqref{eq:PHS_Model:PHS_states} and $\sqnm$ as a coupling input for the third state in \eqref{eq:PHS_Model:PHS_states}, the \ac{PHS} models in \eqref{eq:PHS_Model:Split_RL} and \eqref{eq:PHS_Model:Split_C} are obtained directly from \eqref{eq:PHS_Model:Nonlinear_PHS_full}.
\end{pf}
Splitting the gas pipeline into separate \acp{PHS} more clearly demonstrates the electrical analogy depicted in \autoref{fig:Prelim:Pipeline} and allows for additional properties to be established, as discussed in the following remarks.
\begin{remark}[Monotonic damping] \label{rem:PHS_Model:Monotonic_Damping}
	Despite the complexity of laminar and turbulent friction, the general characteristics are well known (see e.g. the well-known Moody diagram). Specifically, $\sfricL$ \eqref{eq:Prelim:Friction_laminar} decreases linearly and $\sfricT$ \eqref{eq:Prelim:Friction_turbulent_implicit} decreases monotonically in a convex manner w.r.t. $\sqnabs$. The largest rate of change of $\sfrice$ w.r.t. $\sqnabs$ is thus either in the laminar flow region or at the border to the turbulent flow region. From the rate of change of $\sfrice$ at these points, it can be verified that the resistive structures $\mRnx$ in \eqref{eq:PHS_Model:R_Matrix} and $\sRnx[\denoteMid]$ in \eqref{eq:PHS_Model:Split_RL} increase monotonically w.r.t.\ $\sqnabs$. This is valid even if \autoref{asm:PHS_Model:Constant_values} does not hold, since $\scsqr > 0$.
\end{remark}
\begin{remark}(\Ac{OFP} pipelines) \label{rem:PHS_Model:OFP_RL_Pipeline}
	Since the resistive structure $\sRnx[\denoteMid]$ in \eqref{eq:PHS_Model:Split_RL} increases monotonically w.r.t.\ $\sqnabs$, a lower bound for $\sRnx[\denoteMid]$ can be found by considering the laminar flow case where $\sqnabs$ is small. Combining \eqref{eq:Prelim:Reynolds}, \eqref{eq:Prelim:Friction_laminar}, and \eqref{eq:Prelim:Friction_efficiency} for laminar flow yields
	\begin{equation} \label{eq:PHS_Model:Laminar_damping}
		\sRn[\denoteMid] = \frac{32 \srhon \scsqr \svisc}{\sfricEff^2 \sdiam^2 \spMean}, \quad \text{if} \; \sRe < 2300,
	\end{equation}
	which is independent of $\sqnabs$. Using these results, the resistive-inductive \ac{PHS} dynamics of the pipeline in \eqref{eq:PHS_Model:Split_RL} can easily be shown to be \ac{OFP} with a passivity index given by \eqref{eq:PHS_Model:Laminar_damping}. This \ac{OFP} property can for example be used in the analysis and design of interconnected passive systems as in \cite{Malan2022b}.
\end{remark}
\begin{remark}(\Acl{EIP} pipelines) \label{rem:PHS_Model:EIP}
	\break The \acp{PHS} in \autoref{thm:PHS_Model:PHS_Pipe} and \eqref{eq:PHS_Model:Split_RL} can be shown to be \ac{EIP} if the inclination is zero, i.e. $\sinsinc = 0$. Moreover, the \ac{PHS} for the capacitive effects of the pipeline in \eqref{eq:PHS_Model:Split_C} is linear and thus \ac{EIP}.
\end{remark}
\begin{remark}[Necessity of \autoref{asm:PHS_Model:Constant_values}] \label{rem:PHS_Model:Requirement_Z_const}
	\autoref{asm:PHS_Model:Constant_values} allows the complex implicit feedback between $\spp$, $\sZ$ and $\scsqr$ to be neglected. This greatly simplifies the construction of the \ac{PHS} in \eqref{eq:PHS_Model:Nonlinear_PHS}, specifically the choice of $\vx$, without a significant loss of model fidelity, as demonstrated in the sequel. If \autoref{asm:PHS_Model:Constant_values} does not hold, the storage weights for the capacitive terms in $\mQ$ \eqref{eq:PHS_Model:Weight_matrix} and $\sQ[i]$ \eqref{eq:PHS_Model:Split_C} become state dependent. In this case, these dependencies would need to be accounted for in $\vxdot$ in \eqref{eq:PHS_Model:Nonlinear_PHS} and $\sxdot[i]$ in \eqref{eq:PHS_Model:Split_C}.
\end{remark}
\begin{remark}[Necessity of \autoref{asm:PHS_Model:Constant_graviational_effect}] \label{rem:PHS_Model:Requirement_pM_const}
	If \autoref{asm:PHS_Model:Constant_graviational_effect} holds, a feedback effect between the pressures $\spl$, $\spr$, and $\spMean$ on the one hand and the flow rate $\sqnm$ on the other can be neglected. Although we show in the sequel that this assumption does not significantly affect the model fidelity, this does not provide a theoretical stability assurance. Thus, in \autorefapp{app:Stab_mean_pressures}, the stability implications of the feedback neglected by \autoref{asm:PHS_Model:Constant_graviational_effect} is investigated.
\end{remark}
\begin{remark}[Gas mixtures and loads] \label{rem:PHS_Model:Gas_types}
	The standard models in \autoref{sec:Prelim} implicitly assume homogenous gas mixtures. Still, these equations along with the proposed \ac{PHS} models may be used for various homogenous gas compositions by appropriately adjusting the gas properties\footnote{Specifically $\svisc$, $\sR$, $\sTc$, $\spc$ and $\srhon$.} (see e.g.\ \cite{Wiid2020}). Note however, that gas mixtures may exhibit different calorific values. Since end-users typically need a certain power in $\si{\kilo\watt}$, the required volumetric flow rates will change depending on the gas mixture. We also highlight the similarities between constant power gas loads using the volumetric flow rate under standard conditions and constant current electrical loads.
\end{remark}
\begin{remark}[Hydraulic models] \label{rem:PHS_Model:Hydraulic_models}
	The equations and models proposed in this paper may also be used to describe the flow of liquids. To achieve this, replace the relation between pressure $\spp$ and density $\srho$ in \eqref{eq:Prelim:Real_gas_law_isothermal} with the isothermal bulk modulus
	\begin{equation} \label{eq:PHS_Model:Buld_modulus}
		\sK = -\sVol \, \dPartial{\spp}{\sVol} = \srho \dPartial{\spp}{\srho},
	\end{equation}
	with the volume $\sVol = \sm/\srho$ and the mass $\sm$. Substituting 
	\begin{equation} \label{eq:PHS_Model:Convert_to_pressure}
		\dPartial{\spp}{\st} = \dPartial{\spp}{\srho} \dPartial{\srho}{\st},
	\end{equation}
	along with \eqref{eq:Prelim:Flow_rate_relation} and \eqref{eq:PHS_Model:Buld_modulus} into \eqref{eq:Prelim:Pipe_mass_conserv} leads to the simplified momentum \ac{PDE}
	\begin{equation} \label{eq:PHS_Model:hydraulic_momentum}
		\frac{\srho}{\sK} \dPartial{\spp}{\st} = - \frac{\srhon}{\sarea} \dPartial{\sqn}{\sll} .
	\end{equation}
	For incompressible liquids with $\sK \to \infty$, the left-hand side of \eqref{eq:PHS_Model:hydraulic_momentum} becomes zero, yielding the hydraulic equations used e.g.\ in district heating networks in \cite{Strehle2022}.
\end{remark}
\subsection{Network Description} \label{sec:PHS_Model:Network}
Building on the results in \autoref{sec:PHS_Mode:PHS}, we now construct a \ac{PHS} model for a network of gas pipelines. Consider a network $\graphG$ as in \autoref{fig:Simulation:network}, where the edges $\graphE$ represent gas pipelines and the vertices $\graphV$ are points where gas is injected or extracted from the network. Note that for a given node $i \in \graphV$ which describes a pressure $\spp[i]$ connecting to several pipelines $ij \in \graphE$, the capacitive dynamics in \eqref{eq:PHS_Model:Split_C} can be added together to find the equivalent capacitance
\begin{equation} \label{eq:PHS_Model:Node_capacitance}
	\sCeq[i] = \vbT[i]\Diag\left[\frac{\slen[ij]\sarea[ij]}{2\srhon\scsqr}\right]\vb[i],
\end{equation}
where $\mB = (\vbT[i])$ is the incidence matrix of $\graphG$.
\begin{theorem}[Gas network \ac{PHS} model] \label{thm:PHS_Model:Network_PHS}
	Consider a graph $\graphG$ comprising $|\graphV|$ nodes interconnected by $|\graphE|$ pipelines, where $\sqn[i]$ describes the gas injected ($>0$) or extracted ($<0$) at a node $i \in \graphV$. Let \autorefMulti{asm:PHS_Model:Constant_values, asm:PHS_Model:Positive_pressure, asm:PHS_Model:Constant_graviational_effect} hold for the pipelines. Then, the network dynamics can be written as the \ac{PHS}
	\begin{subequations} \label{eq:PHS_Model:Network_PHS_full}
		\begin{equation} \label{eq:PHS_Model:Network_PHS}
			\left\lbrace
			\begin{aligned}
				\vxdot &= (\mJ - \mRnx)\vdHdx + \mG \vu + \mE \vd, \\
				\vy &= \mGT \vdHdx , \quad
				\vz = \mET \vdHdx , \\
				\sHx &= \frac{1}{2} \vxT \mQ \vx,
			\end{aligned}
			\right.
		\end{equation}
		with states \eqref{eq:PHS_Model:Network_PHS_states}, co-states \eqref{eq:PHS_Model:Network_PHS_co_states}, input-output port pair \eqref{eq:PHS_Model:Network_PHS_inputs_outputs}, and disturbance port pair \eqref{eq:PHS_Model:Network_PHS_disturbance}
		\begingroup
		\allowdisplaybreaks
		\begin{align}
			\label{eq:PHS_Model:Network_PHS_states}
			\vx &= \begin{bmatrix}
				\left(\sCeq[i] \spp[i]\right)^\Transpose \;\;& \left(\dfrac{\srhon \slen[ij]}{\sarea[ij]} \sqnm[ij]\right)^\Transpose
			\end{bmatrix}^\Transpose, \\
			\label{eq:PHS_Model:Network_PHS_co_states}
			\vdHdx &= \mQ\vx = \begin{bmatrix}
				\vp \\ \vqnm
			\end{bmatrix},\\
			\label{eq:PHS_Model:Network_PHS_inputs_outputs}
			\vu &= (\sqn[i]),
			\qquad
			\vy = (\spp[i]),\\
			\label{eq:PHS_Model:Network_PHS_disturbance}
			\vd &= \left(\frac{\sg \slen[ij] \sinsinc[ij]}{\scsqr} \spMean[ij]\right),
			\qquad
			\vz = (\sqnm[i]) ,
		\end{align}
		\endgroup
		where $\vd$ is constant and with the interconnection structure \eqref{eq:PHS_Model:Network_J_matrix}, resistive structure \eqref{eq:PHS_Model:Network_R_Matrix}, input matrix \eqref{eq:PHS_Model:Network_Input_matrix}, disturbance matrix \eqref{eq:PHS_Model:Network_Disturbance_matrix} and storage matrix \eqref{eq:PHS_Model:Network_Weight_matrix}
		\begingroup
		\allowdisplaybreaks
		\begin{align}
			\label{eq:PHS_Model:Network_J_matrix}
			\mJ &= \begin{bmatrix}
				\vec{0} & -\mB \\
				\mBT & \vec{0}
			\end{bmatrix},\\
			\label{eq:PHS_Model:Network_R_Matrix}
			\mRnx &= \Diag\left[\vec{0}_{|\graphV|}, \; \left(\frac{\sfrice[ij] \srhon \scsqr \slen[ij] \sqnmabs[ij]}{2 \sdiam[ij] \sarea[ij]^2 \spMean[ij]}\right)\right],\\
			\label{eq:PHS_Model:Network_Input_matrix}
			\mG &= \Diag\left[\Ident[|\graphV|], \; \vec{0}_{|\graphE|}\right],\\
			\label{eq:PHS_Model:Network_Disturbance_matrix}
			\mE &= \Diag\left[\vec{0}_{|\graphV|}, \; -\Ident[|\graphE|]\right],\\
			\label{eq:PHS_Model:Network_Weight_matrix}
			\mQ &= \Diag\left[ \left(\sCeqinv[i]\right), \; \left(\frac{\sarea[ij]}{\srhon \slen[ij]}\right) \right].
		\end{align}
		\endgroup
	\end{subequations}
\end{theorem}
\begin{pf}
	Consider several pipelines connecting to the same node $i \in \graphV$, where one side of each pipeline has the capacitive dynamics described by \eqref{eq:PHS_Model:Split_C}. Since these dynamics all act on the same pressure variable $\spp[i]$, the combined dynamics at node $i$ results in 
	\begin{equation} \label{eq:PHS_Model:Node_dynamics}
		\sCeq[i] \spdot[i] = \sqn[i] - \vbT[i]\vqnm, \quad i \in \graphV,
	\end{equation}
	with $\sCeq[i]$ as in \eqref{eq:PHS_Model:Node_capacitance}, $\vqnm = (\sqnm[ij])$, and $\cramped{\mB = (\vbT[i])}$ the incidence matrix of $\graphG$. The edges of $\graphG$ then comprise the remaining inductive-resistive components from \eqref{eq:PHS_Model:Split_RL}, i.e.\
	\begin{equation} \label{eq:PHS_Model:Edge_dynamics}
		\begin{aligned}
			\frac{\srhon \slen[ij]}{\sarea[ij]} \sqnmdot[ij] =& {-} \frac{\sfrice[ij] \srhon^2 \scsqr \slen[ij] \sqnmabs[ij]}{2 \sdiam[ij] \sarea[ij]^2 \spMean[ij]} \sqnm[ij] + \vbT[ij] \vp \\ &- \frac{\sg \slen[ij] \sinsinc[ij]}{\scsqr} \spMean[ij], \qquad ij \in \graphE
		\end{aligned}
	\end{equation}
	with $\vp = (\spp[i])$, and where $\mBT = (\vbT[ij])$. Combining the vector forms of \eqref{eq:PHS_Model:Node_dynamics} and \eqref{eq:PHS_Model:Edge_dynamics} yields the \ac{PHS} in \eqref{eq:PHS_Model:Network_PHS_full}.
	\qedsymbol
\end{pf}
\autoref{thm:PHS_Model:Network_PHS} allows an entire gas network to be described as a \ac{PHS}. Note the similarity between this resulting \ac{PHS} and, for example, the network description of DC microgrids (see \cite{Strehle2020DC}), which may be exploited for transferring existing control and analysis methods.
\begin{remark}(Supply nodes) \label{rem:PHS_Model:Constant_pressure}
	A supply node designating a fixed pressure $\spp[i]$ can also be included in the network \ac{PHS} \eqref{eq:PHS_Model:Network_PHS_full} by setting $\spdot[i] = 0$ in \eqref{eq:PHS_Model:Node_dynamics} for this node. This eliminates a state in \eqref{eq:PHS_Model:Network_PHS_states} and an input $\sqn[i]$ in \eqref{eq:PHS_Model:Network_PHS_inputs_outputs} and instead treats $\spp[i]$ as a new input which acts on \eqref{eq:PHS_Model:Edge_dynamics}.
\end{remark}

	\section{Simulation} \label{sec:Simulation}
\begin{figure}[!t]
	\centering
	\resizebox{0.7\columnwidth}{!}{%
		\tikzsetnextfilename{03_Img/network_simulation}%
\begin{tikzpicture}
	\def\nodeXdist{5cm}
	\def\nodeYdist{2.4cm}
	\def\nodeXshift{\nodeXdist/2}
	
	\def\busGendist{20pt}
	
	\def\Nodes{1,2,3}
	\def\NodesGeneration{
		1/east/1*/0*/$\spp[1]$/latex'-,
		2/west/-1*/0*/$\sqn[2]$/-latex',
		3/north/0*/1*/$\sqn[3]$/-latex'
	}
	
	\tikzset{noMiddle segment/.style={decoration={noMiddle},decorate, segment length=#1}}
	
	\tikzstyle{bus}				= [draw, circle, inner sep=3pt, fill=black!20!white]
	\tikzstyle{generation}		= [draw, circle, inner sep=1pt, fill=black!5!white]
	
	\tikzstyle{linePhysPerm}	= [-, line width=1.0pt]
	
	\coordinate(cNode1);
	\path (cNode1) ++(0.4*\nodeXdist,\nodeYdist) coordinate (cNode2);
	\path (cNode1) ++(-0.8*\nodeXdist,0) coordinate (cNode3);
	
	\node[bus](node1) at(cNode1) {1};
	\node[bus](node2) at(cNode2) {2};
	\node[bus](node3) at(cNode3) {3};

	\foreach \a/\b/\c/\d/\e/\f in \NodesGeneration {
		\path (node\a.\b) ++(\c\busGendist,\d\busGendist) coordinate (cGen\a);
		\node[generation](gen\a) at(cGen\a) {\small \e};
		\draw[\f, line width=1.0pt] (node\a.\b) -- (gen\a);
	}
	
	\draw[linePhysPerm](node1) -- (node2) node[pos=0.5, anchor=north west, align = left] {\small Line$_{12}$,\\\small$\slen[12]=\SI{80}{\kilo\meter}$};
	\draw[linePhysPerm](node1) -- (node3) node[pos=0.5, anchor=north, sloped] {\small Line$_{13}$, $\slen[13]=\SI{90}{\kilo\meter}$};
	\draw[linePhysPerm](node2) -- (node3) node[pos=0.5, anchor=south, sloped] {\small Line$_{23}$, $\slen[23]=\SI{100}{\kilo\meter}$};
\end{tikzpicture}
%
	}
	\vspace*{-10pt}
	\caption{Gas network comprising three nodes.}
	\label{fig:Simulation:network}
\end{figure}
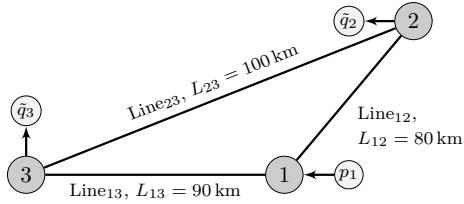

We now demonstrate the model fidelity of the lumped-parameter model in \autorefMulti{prop:PHS_Model:Mass_ODE, prop:PHS_Model:Momentum_ODE} and the proposed \ac{PHS} model in \autoref{thm:PHS_Model:PHS_Pipe} with a \Matlab/\Simscape simulation of the benchmark three-node network in \autoref{fig:Simulation:network}. Furthermore, the obtained results are compared with the results from \cite{Ke2000,HerranGonzalez2009,Alamian2012,Pambour2016}, and the effects of \autorefMulti{asm:PHS_Model:Constant_values, asm:PHS_Model:Positive_pressure, asm:PHS_Model:Constant_graviational_effect} are investigated. 
\subsection{Simulation Setup} \label{sec:Simulation:Setup}
%
%
\begin{table}[!t]
	\centering
	\captionsetup{width=.9\columnwidth}
	\caption{Simulation Parameter Values}
	\label{tab:Simulation:Sim_params}
	\renewcommand{\arraystretch}{1.25}
	\begin{tabular}{lcc}
		\noalign{\hrule height 1.0pt}
		Parameter & Symbol & Value \\
		\noalign{\hrule height 1.0pt}
		Specific gas constant & $\sR$ & $\SI{518.28}{\joule\per\kilogram\per\kelvin}$ \\
		Dynamic viscosity & $\svisc$ & $\SI{e-5}{\kilogram\meter\per\second}$ \\
		Critical pressure & $\spc$ & $\SI{46.5}{\bar}$ \\
		Standard pressure & $\spn$ & $\SI{1.01325}{\bar}$ \\
		Critical temperature & $\sTc$ & $\SI{190.55}{\kelvin}$ \\
		Standard temperature & $\sTn$ & $\SI{273.15}{\kelvin}$ \\
		Simulation temperature & $\sT$ & $\SI{278}{\kelvin}$ \\
		\hline
		Friction efficiency factor & $\sfricEff$ & $0.98$ \\
		Pipe roughness & $\srough$ & $\SI{0.012}{\milli\meter}$ \\
		Pipe diameter & $\sdiam$ & $\SI{0.6}{\meter}$ \\
		\noalign{\hrule height 1.0pt}
	\end{tabular}
\end{table}
\begin{figure}[!t]
	\centering
	\resizebox{\columnwidth}{!}{\includegraphics[scale=1]{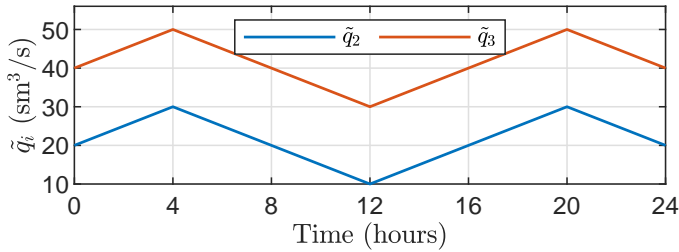}}
	\vspace*{-18pt}
	\caption{Flow rate of the loads at Nodes~2 and 3.}
	\label{fig:Simulation:loads}
\end{figure}
The gas network in \autoref{fig:Simulation:network} is simulated using the parameter values in \autoref{tab:Simulation:Sim_params}. The pressure at Node~1 is kept constant at $\spp[1] = \SI{50}{\bar}$ and the loads at Nodes 2 and 3 are set as in \autoref{fig:Simulation:loads}.
Furthermore, to investigate the model fidelity in the presence of non-zero inclination angles, the elevation of Node~1 relative to Nodes~2 and 3 is changed to one of the heights $\sh[1] \in \lbrace{-1,-0.5,0.5,1}\rbrace\si{\kilo\meter}$, where $\slen[1j]\sinsinc[1j] = \sh[j] - \sh[1]$ with $1j \in \graphE$. Note that $\sinc[23] = 0$.
\subsection{Results} \label{sec:Simulation:Reults}
\begin{figure*}[!ht]
	\centering
	\vspace*{4pt}
	\begin{minipage}[t!]{\textwidth}
		\resizebox{\textwidth}{!}{\includegraphics[scale=1]{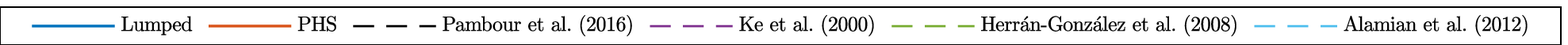}}
		\vspace*{-4pt}
	\end{minipage}
	\begin{minipage}[t!]{0.45\textwidth}
		\centering
		\resizebox{\textwidth}{!}{\includegraphics[scale=1]{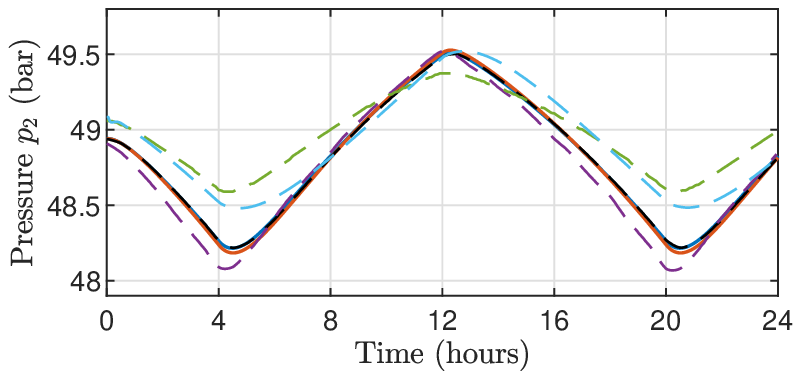}}
		\vspace*{-18pt}
		\caption{Pressures at Node~2.}
		\label{fig:Simulation:Pressures_P2}
	\end{minipage}
	\hspace*{22pt}
	\begin{minipage}[t!]{0.45\textwidth}
		\centering
		\resizebox{\textwidth}{!}{\includegraphics[scale=1]{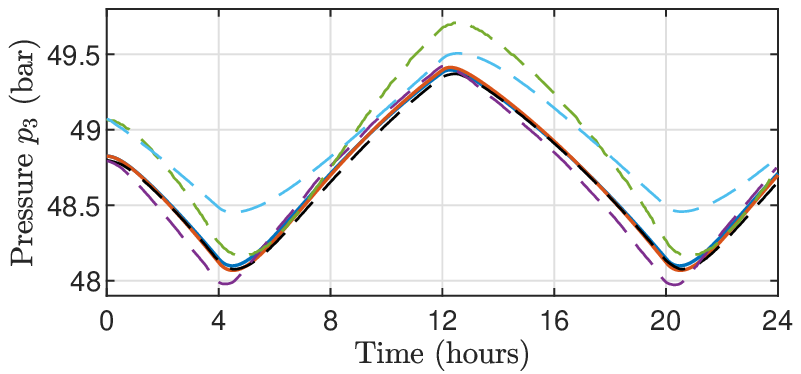}}
		\vspace*{-18pt}
		\caption{Pressures at Node~3.}
		\label{fig:Simulation:Pressures_P3}
	\end{minipage}
\end{figure*}
\begin{figure}[!t]
	\centering
	\resizebox{\columnwidth}{!}{\includegraphics[scale=1]{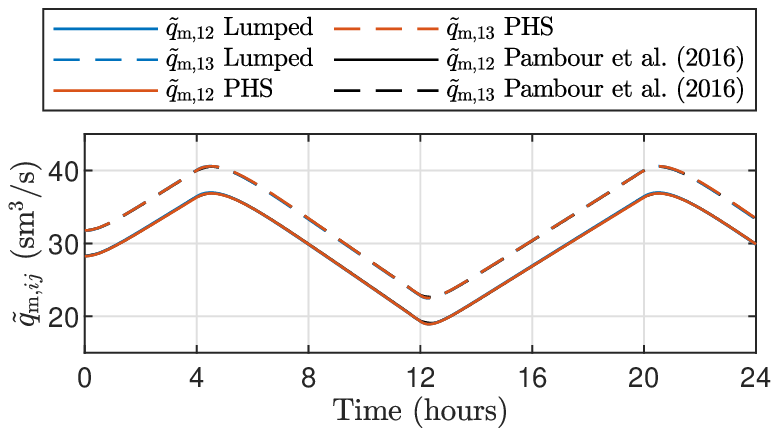}}
	\vspace*{-24pt}
	\caption{Volumetric flow rates at standard conditions in Pipeline$_{12}$ and Pipeline$_{13}$.}
	\label{fig:Simulation:Flowrates}
\end{figure}
\begin{figure}[!t]
	\centering
	\resizebox{\columnwidth}{!}{\includegraphics[scale=1]{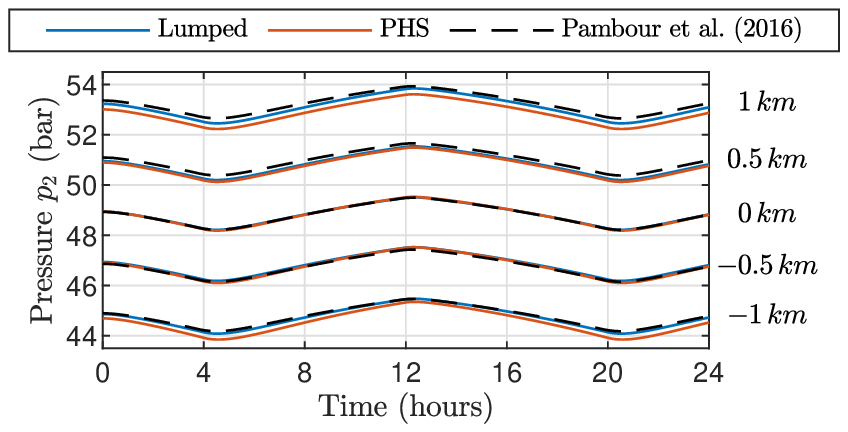}}
	\vspace*{-18pt}
	\caption{Pressure at Node~2 with the respective elevations of Node~1 given on the right.}
	\label{fig:Simulation:Heights_P2}
\end{figure}
\begin{figure}[!t]
	\centering
	\resizebox{\columnwidth}{!}{\includegraphics[scale=1]{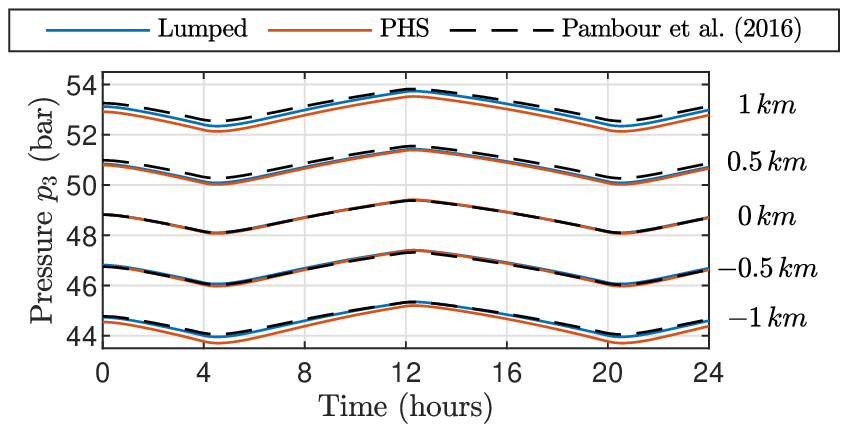}}
	\vspace*{-18pt}
	\caption{Pressure at Node~3 with the respective elevations of Node~1 given on the right.}
	\label{fig:Simulation:Heights_P3}
\end{figure}
The simulated pressures at Nodes~2 and 3 are shown in \autoref{fig:Simulation:Pressures_P2} and \autoref{fig:Simulation:Pressures_P3}, respectively, while volumetric flow rates $\sqnm[12]$ and $\sqnm[13]$ are shown in \autoref{fig:Simulation:Flowrates}. Furthermore, the pressures at $\spp[2]$ and $\spp[3]$ with Node~1 at various elevations are shown \autoref{fig:Simulation:Heights_P2} and \autoref{fig:Simulation:Heights_P3}, respectively. In each case, results are shown for the lumped-parameter model comprising \autorefMulti{prop:PHS_Model:Mass_ODE, prop:PHS_Model:Momentum_ODE} and for the \ac{PHS} model in \autoref{thm:PHS_Model:PHS_Pipe} where \autorefMulti{asm:PHS_Model:Constant_values, asm:PHS_Model:Positive_pressure, asm:PHS_Model:Constant_graviational_effect} are applied.

We note the high level of similarity of both models compared to the results in \cite{Pambour2016}\footnote{We use \cite{Pambour2016} as a benchmark due to its high accuracy compared to the commercial simulation software SIMONE.}. The pressures in both models show deviations of no more than $\SI{0.13}{\percent}$ and the volumetric flow rates of both models exhibit deviations of no more than $\SI{1.05}{\percent}$. Without elevation differences in the network, there is no significant difference in quality between the lumped-parameter and \ac{PHS} models. However, the effects of \autorefMulti{asm:PHS_Model:Constant_values, asm:PHS_Model:Positive_pressure, asm:PHS_Model:Constant_graviational_effect} show more clearly when elevation differences are present. The simulated pressures in \autoref{fig:Simulation:Heights_P2} and \autoref{fig:Simulation:Heights_P3} show maximal deviations of $\SI{0.39}{\percent}$ for the lumped-parameter model and $\SI{0.81}{\percent}$ for the \ac{PHS} model. Nevertheless, these errors can be considered sufficiently small for models aimed towards analysis and design.

	\section{Conclusion} \label{sec:Conclusion}
In this paper, we proposed a third-order \ac{PHS} model for gas pipelines based on the discretized and simplified Euler equations.
We showed that this model exhibits the same structure as $\pi$-model transmission lines, can be combined into a \ac{PHS} model for a gas network, and has passivity properties conducive to passivity-based control.
Furthermore, the simulation results demonstrate a model fidelity comparable to that of detailed simulation models.
Future work includes exploiting the parallels drawn between the gas and power networks and using the the \ac{PHS} framework for a conjoined consideration of networked multi-energy systems.

	
	\appendix
	\section{Stability for variable mean pressures} \label{app:Stab_mean_pressures}
We here investigate the stability of the feedback effect as described by \autoref{rem:PHS_Model:Requirement_pM_const}. Consider therefore the model in \autoref{thm:PHS_Model:PHS_Pipe} with a non-constant average pressure $\spMean$, i.e.\ \autoref{asm:PHS_Model:Constant_graviational_effect} does not hold. Rearranging \eqref{eq:PHS_Model:Mean_Pressure} results in
\begin{equation} \label{eq:PHS_Model:Mean_Pressure_modified}
	\spMean = \underbrace{\frac{1}{3} \left(2 - \frac{\spr}{\spl + \spr}\right)}_{\textstyle \skl} \spl + \underbrace{\frac{1}{3} \left(2 - \frac{\spl}{\spl + \spr}\right)}_{\textstyle \skr} \spr ,
\end{equation}
where $\skl, \skr \in (\nicefrac{1}{3},\nicefrac{2}{3})$ if $\spl > 0$ and $\spr > 0$. Substituting \eqref{eq:PHS_Model:Mean_Pressure_modified} into the \ac{PHS} pipeline dynamics \eqref{eq:PHS_Model:Nonlinear_PHS} yields
\begin{equation} \label{eq:PHS_Model:var_mean_pressure_system}
	\mQ\!\!\begin{bmatrix} \spldot \\ \sprdot \\ \sqnmdot \end{bmatrix} \!=\,
	\underbrace{\!\!\begin{bmatrix}
			0 & 0 & -1 \\
			0 & 0 & 1 \\
			1 {-} \sphi\skl & -1 {-} \sphi\skr & -\sRn[\denoteMid](\sqnmabs, \spMean)
		\end{bmatrix}\!\!}_{\mA} \!
	\begin{bmatrix}
		\spl \\ \spr \\ \sqnm
	\end{bmatrix} + 
	\begin{bmatrix}
		\sqnl \\ \sqnr \\ 0
	\end{bmatrix}\!,
\end{equation}
where $\sphi \coloneqq \sg \slen \sinsinc / \scsqr$ and with $\mQ$ in \eqref{eq:PHS_Model:Weight_matrix}. Note that \eqref{eq:PHS_Model:var_mean_pressure_system} is no longer a \ac{PHS}, since the symmetric part of the state matrix $\mA$ is no longer positive semi-definite.
\begin{proposition} \label{thm:PHS_Model:Stable_var_mean_pressure}
	The gas pipeline dynamics with a variable mean pressure described by \eqref{eq:PHS_Model:var_mean_pressure_system} are Lyapunov stable if 
	\begin{equation} \label{eq:PHS_Model:stability_height_diff}
		\slen\sinsinc < \frac{6\scsqr}{\sg}.
	\end{equation}
\end{proposition}
\begin{pf}
	Since the constant matrix $\mQ$ is positive definite, the Lyapunov stability of \eqref{eq:PHS_Model:var_mean_pressure_system} can be evaluated by looking at the eigenvalues of the state matrix $\mA$. Specifically, the solution of \eqref{eq:PHS_Model:var_mean_pressure_system} is guaranteed to be stable if the eigenvalues of $\mA$ are nonpositive everywhere along the state trajectory. The eigenvalues of $\mA$ are
	\begin{equation} \label{eq:PHS_Model:Eigenvalues_A}
		\seig[1] \!\;{=}\!\; 0, \;\; \seig[2,3] \!\;{=} -\frac{1}{2}\sRn[\denoteMid] \pm \frac{1}{2}\sqrt{\sRn[\denoteMid]^2 -\!\!\: 8 \!\!\:- 4 \sphi \skl + 4 \sphi \skr}.
	\end{equation}
	Since $\sRn[\denoteMid] > 0$ (see \autoref{thm:PHS_Model:PHS_Pipe}), positive eigenvalues can only be obtained if the rooted term in \eqref{eq:PHS_Model:Eigenvalues_A} is positive. Evaluating where the eigenvalues in \eqref{eq:PHS_Model:Eigenvalues_A} are negative when the rooted term is positive thus leads to
	\begin{equation} \label{eq:PHS_Model:ineq_first_step}
		\begin{aligned}
			&&\frac{1}{2}\sqrt{\sRn[\denoteMid]^2 - 8 - 4 \sphi \skl + 4 \sphi \skr} &< \frac{1}{2}\sRn[\denoteMid] \\
			\implies&& \sRn[\denoteMid]^2 - 8 - 4 \sphi \skl + 4 \sphi \skr &< \sRn[\denoteMid]^2 \\
			\iff&& \sphi (\skr - \skl) &< 2.
		\end{aligned}
	\end{equation}
	From \eqref{eq:PHS_Model:Mean_Pressure_modified}, we observe that
	\begin{equation} \label{eq:PHS_Model:difference_pressure_gains}
		\frac{\spr - \spl}{\spr + \spl} = 3(\skr - \skl).
	\end{equation}
	Substituting $\sphi = \sg \slen \sinsinc / \scsqr$ and \eqref{eq:PHS_Model:difference_pressure_gains} into \eqref{eq:PHS_Model:ineq_first_step} gives
	\begin{equation} \label{eq:PHS_Model:var_mean_stability_ineq}
		\frac{\spr - \spl}{\spr + \spl}\slen\sinsinc < \frac{6\scsqr}{\sg} .
	\end{equation}
	Since \eqref{eq:PHS_Model:difference_pressure_gains} is bounded by $[-1,1]$, \eqref{eq:PHS_Model:stability_height_diff} is obtained as a sufficient condition for \eqref{eq:PHS_Model:var_mean_stability_ineq}.
	\qedsymbol
\end{pf}
By verifying that \eqref{eq:PHS_Model:stability_height_diff} holds for a given pipeline, \autoref{thm:PHS_Model:Stable_var_mean_pressure} ensures the stability of the underlying system dynamics. This in turn means that no unstable dynamics are hidden away when applying \autoref{asm:PHS_Model:Constant_graviational_effect} to obtain the \ac{PHS} \eqref{eq:PHS_Model:Nonlinear_PHS_full}. Note that the left-hand side of \eqref{eq:PHS_Model:stability_height_diff} is the height difference between the two ends of the pipes and that \eqref{eq:PHS_Model:stability_height_diff} is met in all practical cases\footnote{If $\scsqr=(\SI{300}{\meter\per\second})^2$ and $\sg=\SI{9.805}{\meter\per\second^2}$, the right-hand side of \eqref{eq:PHS_Model:stability_height_diff} evaluates to a difference in height of $\SI{55.07}{\kilo\meter}$.}.
%

	
	\bibliography{ms}

\end{document}